\documentclass{pasj00}
\SetRunningHead{Astronomical Society of Japan}{Usage of \texttt{pasj00.cls}}

\title{ASTE observations of nearby galaxies: A tight correlation between
CO(J=3--2) emission and H$\alpha$
}
\author{Shinya \textsc{Komugi}$^1$,
 Kotaro \textsc{Kohno}$^1$, 
 Tomoka \textsc{Tosaki}$^2$,
 Hiroyuki \textsc{Nakanishi}$^2$,\\
 Sachiko \textsc{Onodera}$^1$,
 Fumi \textsc{Egusa}$^1$
\and Yoshiaki \textsc{Sofue}}

\affil{
$^1$Institute of Astronomy, The University of Tokyo, 2--21--1 Osawa,
Mitaka-shi, Tokyo, 181--8588}

\affil{
$^2$Nobeyama Radio Observatory, Minamisaku, Nagano, 384-1305}
\email{skomugi@ioa.s.u-tokyo.ac.jp}


\Received{$\langle$reception date$\rangle$}
\Accepted{$\langle$acception date$\rangle$}
\Published{$\langle$publication date$\rangle$}
\begin{document}
\maketitle

\begin{abstract}
Star formation rates (SFRs) obtained via extinction corrected H$\alpha$ are
 compared to dense gas as traced by $\atom{CO}{}{12}$ (\textit{J}\ =\
 3--2) emission at the centers of nearby galaxies, observed with the
 ASTE telescope.  It is found that,
 although many of the observed positions are dusty and therefore
 heavily absorbed at H$\alpha$, the SFR shows a striking correlation
 with dense gas in the form of the Schmidt law with an index 1.0.  The
 correlation is also compared between gas traced by  $\atom{CO}{}{12}$ (\textit{J}\ =\
 1--0) and application of H$\alpha$ extinction correction.  We find that
 dense gas produces a far better correlation with SFR in view of surface
 density values.   
\end{abstract}

\section{Introduction}
The present knowledge of star formation on galactic scales in relation with its precursor gas is
generally expressed by the Schmidt law \citep{schmidt},

\begin{equation}
SFR \propto \rho^N
\end{equation}

where $SFR$ is the star formation rate, $\rho$ the
gas density, and $N$ the Schmidt law index,
expressing the efficiency of star formation from gas.  Often written
also in terms of surface averaged quantities, equation (1)
relates two physical values $SFR$ and $\rho$ 
which are generally
spatially decoupled when observed locally, and connected in a spatially
averaged sense.  The connection between the two values also has a time
averaged nature, namely the formation timescale of massive stars.  Therefore, in order to obtain valid physical suggestions
from the Schmidt law, we must derive these two values based on
measurements which express conditions that are spatially and temporally
connected as much as observations allow.  

Previous observations of molecular gas have been conducted mainly in the
 $\atom{CO}{}{12}$ (\textit{J}\ =\ 1--0) line, tracing cold gas which constitutes the
 bulk of galactic molecular clouds.  However, \citet{kohno99} have shown
 that a tracer of denser gas, HCN, shows a better spatial
 correlation with star forming regions.  Observations of \textit{dense} gas tracers such as
 HCN \citep{gao} and higher transition CO lines like $\atom{CO}{}{12}$
 (\textit{J}\ =\ 2--1) \citep{braine, boker} and
 $\atom{CO}{}{12}$ (\textit{J}\ =\ 3--2) \citep{mauersberger, yao, narayanan} are becoming accessible, and it is of interest how
 these tracers of \textit{dense} show up in terms of the Schmidt law.

The calibration of SFR becomes an issue under these circumstances, where virtually all observational inquiry of the
Schmidt law using \textit{dense} gas has resorted to FIR as the SFR
calibrator.  This in part owes to the fact that FIR data from the IRAS
satellite is abundant, and that many of the sample galaxies observed in dense
gas tracers (HCN and higher CO transitions) were selected according to FIR
luminosity.  However, we must bear in mind that the SFR must be
calibrated using \textit{massive} stars.  The SFR derived from FIR
luminosity can only 
trace star formation over $\sim 10^8$ years, because of its
contamination from extended dust heated by the interstellar radiation
field (\cite{K98}), which can amount to a significant fraction of the FIR
luminosity (e.g., \cite{hirashita}).

  This results in overestimation of the
SFR, and also becomes a bottleneck in improving the
temporal connection of gas and star formation.  Another practical shorthand is that the angular
resolution of IRAS data is commonly several arcminutes; too large to
infer physics from the correlation between dense gas tracer data, which
are typically tens of arcseconds in angular resolution. 
 The most reliable massive SFR tracer to date is the
H$\alpha$ luminosity tracing star formation over several $10^6$ years, whose correlation with dense gas, surprisingly has not been
checked.  The main reason is that H$\alpha$ is weak in dusty galaxies
such as those observed in dense gas tracers, due to the extinction
within the galaxies.  It is important therefore not only to acquire the
H$\alpha$ luminosity of these galaxies, but to accurately estimate the amount
of extinction.  

The main objective and result of this Letter is to examine the
correlation between $\atom{CO}{}{12}$ (\textit{J}\ =\ 3--2) tracing
warm dense gas (typically $\sim$ 30K), and
\textit{extinction corrected} H$\alpha$ luminosity tracing accurately the SFR, both in
surface averaged densities.

\section{Observation}
Observation of the $\atom{CO}{}{12}$ (\textit{J}\ =\ 3--2) at 345GHz was
conducted using the Atacama Submillimeter Telescope Experiment (ASTE)
\citep{ezawa04, kohno05}, a 10m single dish
located in the Atacama desert of altitude 4800m in Pampa La Bola,
Chile.  Observations were remotely made
from an ASTE operation room of the National Astronomical Observatory of
Japan (NAOJ) at Mitaka, Japan,
using a network observation system N-COSMOS3 developed
by NAOJ (\cite{kam05}).

The sample was selected to be able to compare the dense gas quantity and its relation with star formation,
 in both normal and starbursting galaxies.  The galaxies were selected
 so that most them have readily accessible and extinction correctable
 H$\alpha$ data.  Another limitation for the samples was that their
 velocity width had to be under 350 $\mathrm{km\ s^{-1}}$, chosen so that the emission
 fits within the bandwidth 445 $\mathrm{km\ s^{-1}}$ of the backend, allowing for
 baseline subtraction.  Table \ref{astesample} lists the observed
 samples.  The galaxies were observed only at their central position,
 and the resolution FWHM of ASTE (22 $^{\prime \prime}$) corresponds to a
 linear size of $\sim $3 kpc for a typical sample distance of 30 Mpc.

\begin{table*}
\caption{Observed Galaxies}\label{astesample}
\begin{center}
\begin{tabular}{lllllllll}  \hline \hline
 Galaxy   & RA   	    & DEC          & Morphology & $D$  & $i$  &
 $\mathrm{I_{CO}^{3-2}}$ & $\log \Sigma_\mathrm{SFR}$ & $\log \Sigma_\mathrm{SFR}^{corr}$     \\
          & 1950 	    & 1950         &            &  Mpc & deg. &
 (K km/s) & $\MO\ \mathrm{pc}^{-2}\ \mathrm{yr}^{-1}$ 
&\\
 (1)      & (2) 	    & (3)          & (4)        & (5)  &  (6) &
 (7)    & (8) & (9)   \\ \hline
NGC\ 157  & $00\ 32\ 14.45$   & $-08\ 40\ 18.8$   & SABbc   & 35   & 49
 & 6.9$\pm$0.5 & -7.58 & \\
NGC\ 520  & $01\ 21\ 59.79^*$ & $+03\ 31\ 55.9^*$ & Pec     & 45.5 & 66
 & no detection & & \\
NGC\ 925  & $02\ 24\ 16.89$   & $+33\ 21\ 18.9$   & SABd    & 14.3 & 53
 & 1.1$\pm$0.5  & -7.70 & \\
NGC\ 1022 & $02\ 36\ 03.99$   & $-06\ 53\ 34.1$   & SBa     & 30.1 & 34
 & 45$\pm$1.0  & & \\
NGC\ 1068 & $02\ 40\ 07.05$   & $-00\ 13\ 31.6$   & SAb     & 22.7 & 40
 & 157$\pm$4.5 & & \\
	  & $02\ 40\ 06.50^*$ & $-00\ 13\ 32.0^*$ &	        &      &
 & 129$\pm$3.8 & -5.42 & -4.92 \\	
NGC\ 1084 & $02\ 43\ 32.11$   & $-07\ 47\ 17.4$   & SAc     & 28.1 & 59
 & 19$\pm$1.7  & & \\
	  & $02\ 43\ 31.80^*$ & $-07\ 47\ 06.0^*$ &         &      &
 & 13.4$\pm$1.0 & -7.51 & \\
NGC\ 1087 & $02\ 43\ 51.88$   & $-00\ 42\ 28.4$   & SABc    & 36.9 & 49
 & 10.3$\pm$0.5  & & \\
	  & $02\ 43\ 51.60^*$ & $-00\ 42\ 19.0^*$ &	        &      &
 & 8.3$\pm$0.2  & -7.17 &\\
NGC\ 7479 & $23\ 02\ 26.39$   & $+12\ 03\ 10.3$   & SBc     & 52.1 & 38
 & 21.3$\pm$0.9 & & \\
	  & $23\ 02\ 26.80^*$ & $+12\ 03\ 06.0^*$ &	        &      &
 & 15.9$\pm$0.8 & -7.77 & -6.74 \\
NGC\ 7625 & $23\ 17\ 60.00$   & $+16\ 57\ 05.6$   & SAa\ pec & 37.3 & 22
 & 24.3$\pm$0.8 & & \\ 
          & $23\ 18\ 00.60^*$ & $+16\ 57\ 15.0^*$ &         &      &
 & 11.3$\pm$0.5 & & \\  \hline \hline
\end{tabular}
\end{center}
Col.(1):Galaxy name.  Col.(2)(3):Coordinates from NED.  Many were
 observed also at optically defined coordinates from \citet{dressel},
 used in the $\atom{CO}{}{12}$ (\textit{J}\ =\ 1--0) survey by Komugi et
 al. (in preparation), marked $^*$.  For NGC\ 157 and
 NGC\ 925, coordinates from both NED and \citet{dressel} match.  Col.(4)(5)(6):Morphology, distance,
 and inclination in degrees, as listed in \citet{young95}.  Col.(7)
 Observed integrated intensity of $\atom{CO}{}{12}$ (\textit{J}\ =\
 3--2), converted to main beam temperature units.
Col.(4)(5): SFR and
 extinction corrected SFR, respectively, as explained in text.
\end{table*}

The observations were conducted in August 2005, in fair weather conditions, using a double sideband cooled SIS mixer.
  Calibration was done using the standard chopper wheel method.  Backend
 was a 1024 channel digital spectrometer with 512 MHz bandwidth and
  frequency resolution of 0.5 MHz, corresponding to a velocity resolution of 0.43$\mathrm{km\ s^{-1}}$. 
 The resulting velocity bandwidth is 445$\mathrm{km\ s^{-1}}$.   Typical
  system temperatures at 345 GHz ranged from 180K to 300K.  Pointing was
  checked every several hours using Uranus or Mars, and was found to be accurate to $\sim$ 1$^{\prime \prime}$.  

Obtained data were reduced with NEWSTAR, an AIPS based software, used
commonly at Nobeyama Radio Observatory (NRO).  After flagging bad spectra, first to
second order baselines were subtracted, then smoothed to a velocity resolution of typically 15 to 20 $\mathrm{km\ s^{-1}}$.  $\mathrm{T_A^*}$ was then 
converted to $\mathrm{T_{mb}}$ via $\mathrm{T_{mb}}=\mathrm{T_A^*}/0.6$,
were 0.6 is the main beam efficiency of the ASTE telescope.  The
resultant integrated intensity of the galaxies are listed in table \ref{astesample}.

\section{Results}
\begin{figure}
  \begin{center}
  \begin{tabular}{cc}
    \FigureFile(40mm,40mm){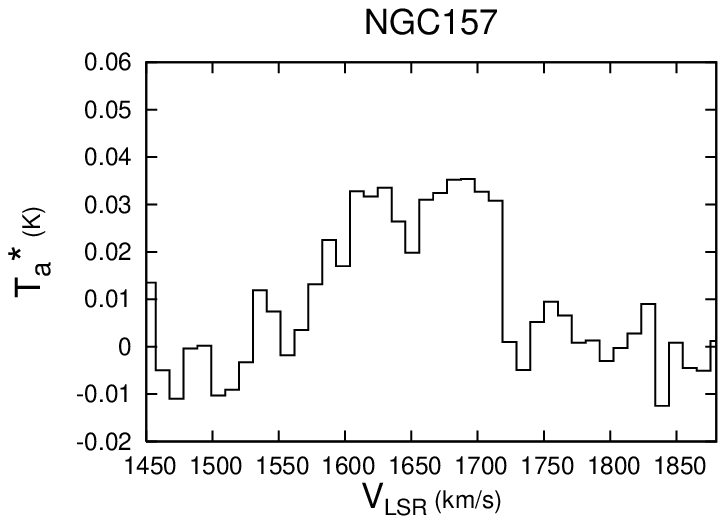}
    \FigureFile(40mm,40mm){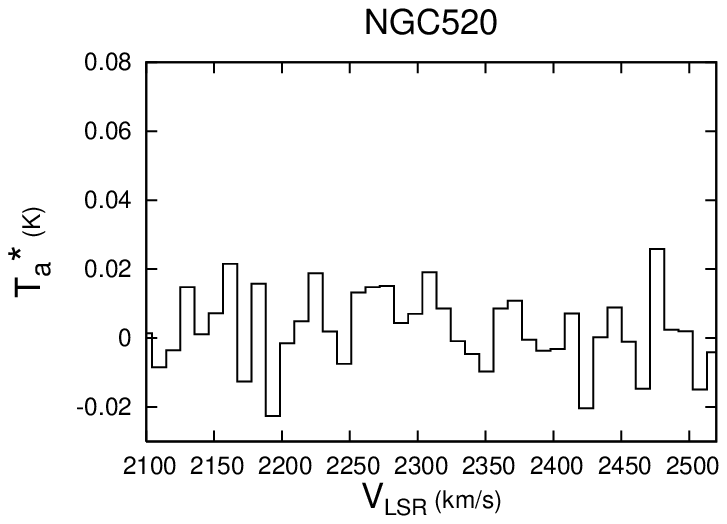}\\
    \FigureFile(40mm,40mm){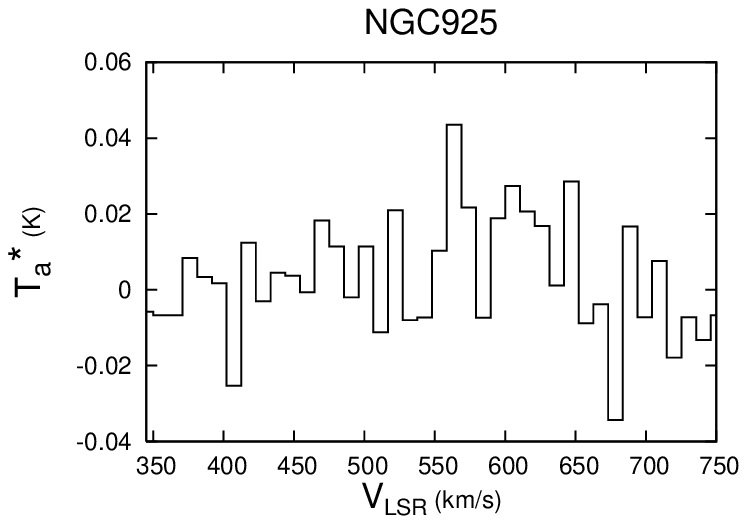}
    \FigureFile(40mm,40mm){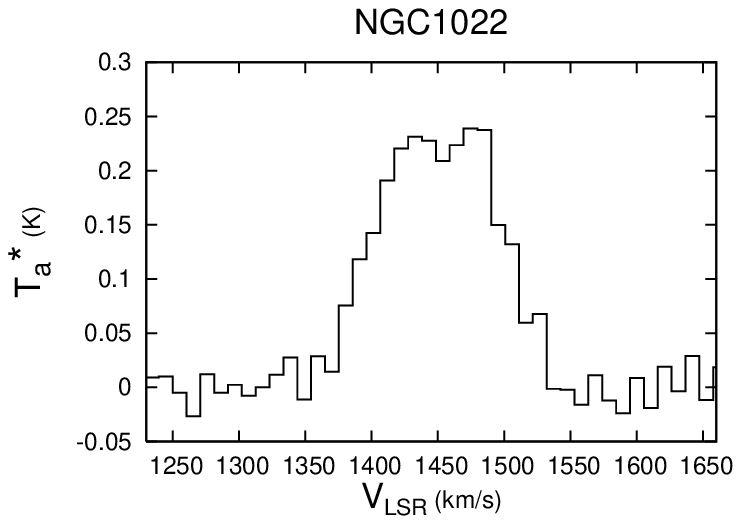}\\
    \FigureFile(40mm,40mm){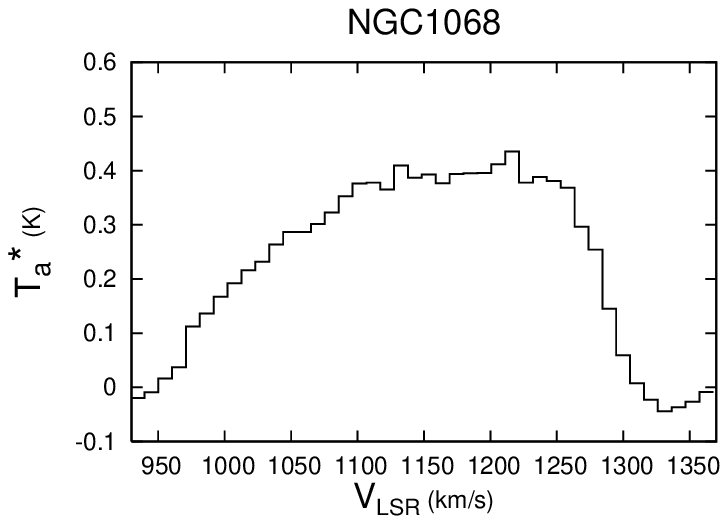}
    \FigureFile(40mm,40mm){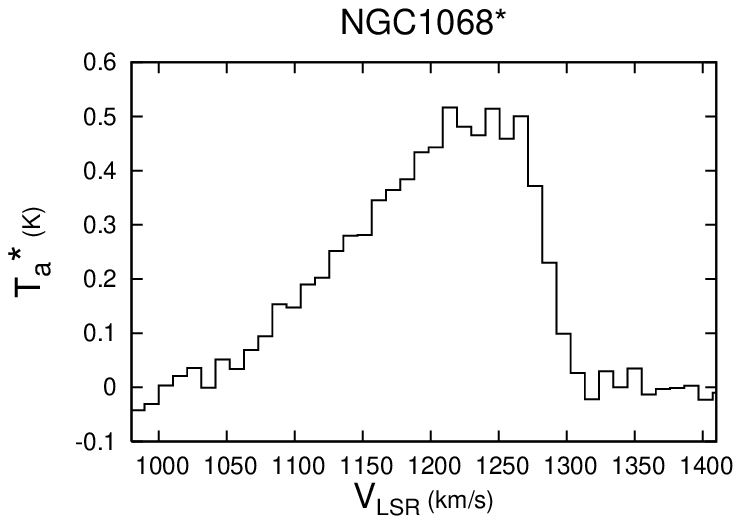}\\
    \FigureFile(40mm,40mm){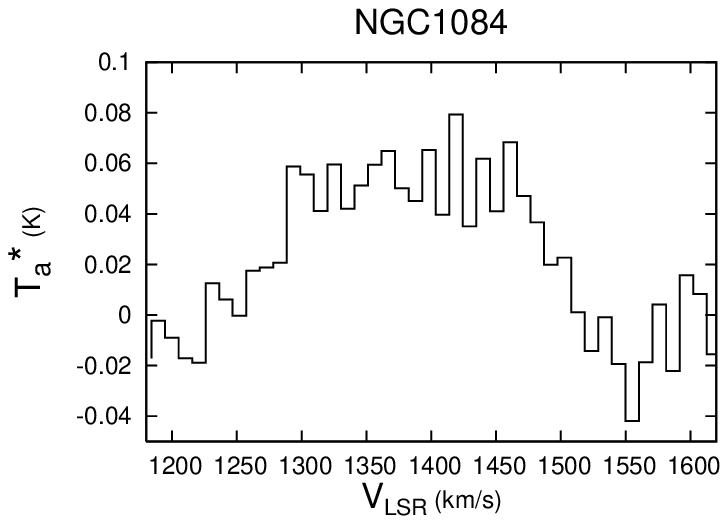}
    \FigureFile(40mm,40mm){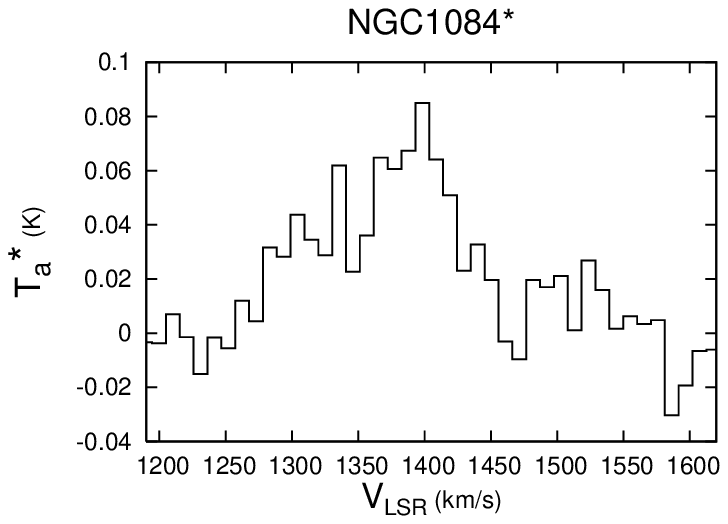}\\
    \FigureFile(40mm,40mm){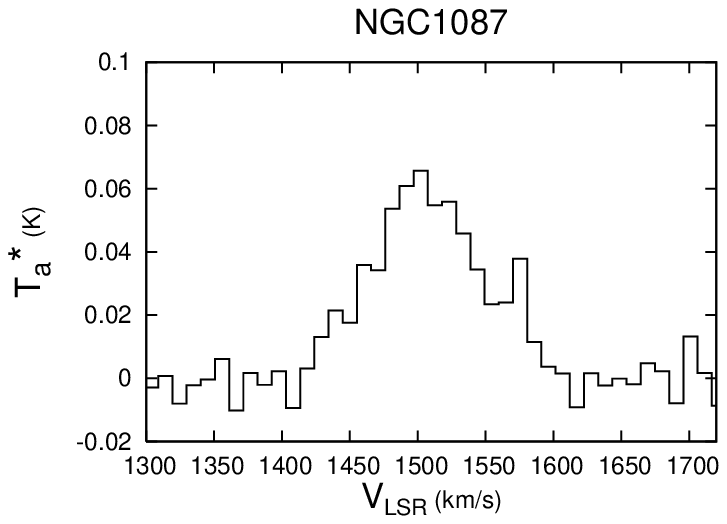}
    \FigureFile(40mm,40mm){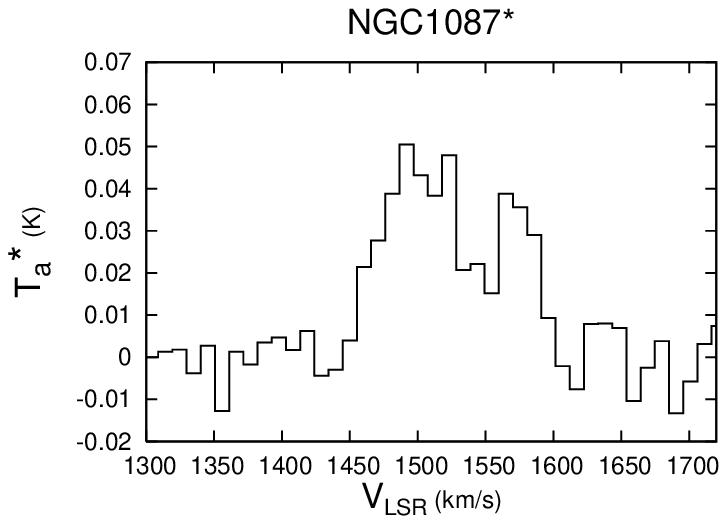}\\
    \FigureFile(40mm,40mm){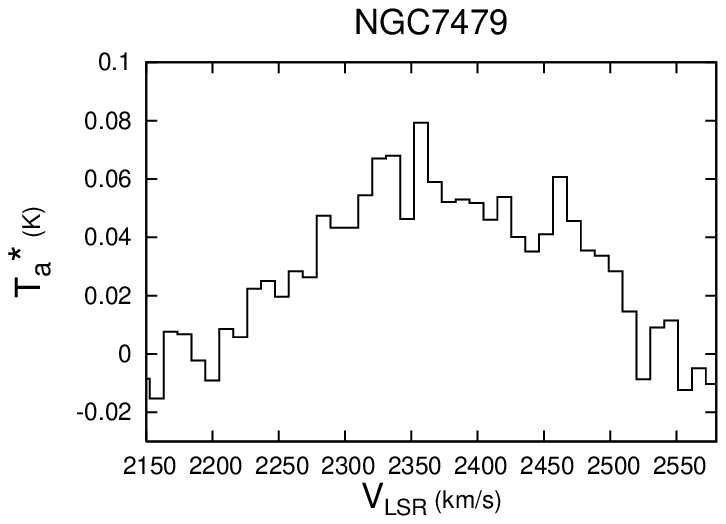}
    \FigureFile(40mm,40mm){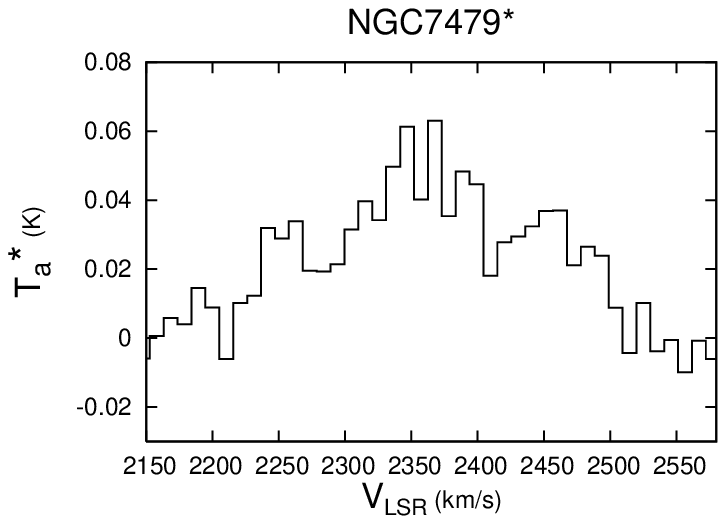}\\
    \FigureFile(40mm,40mm){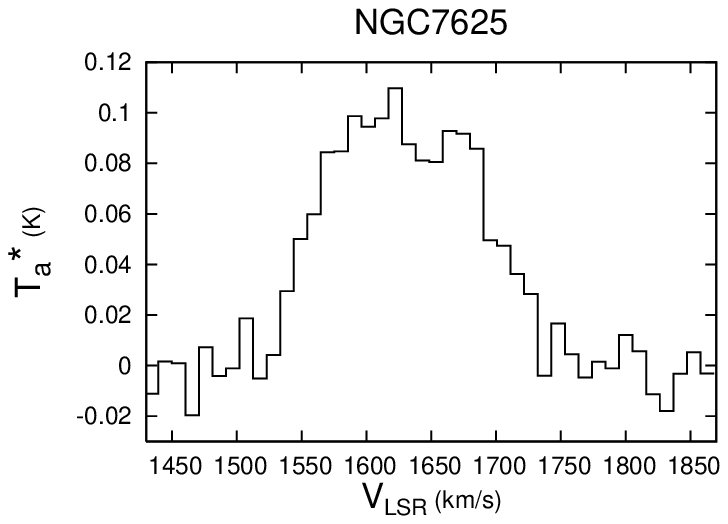}
    \FigureFile(40mm,40mm){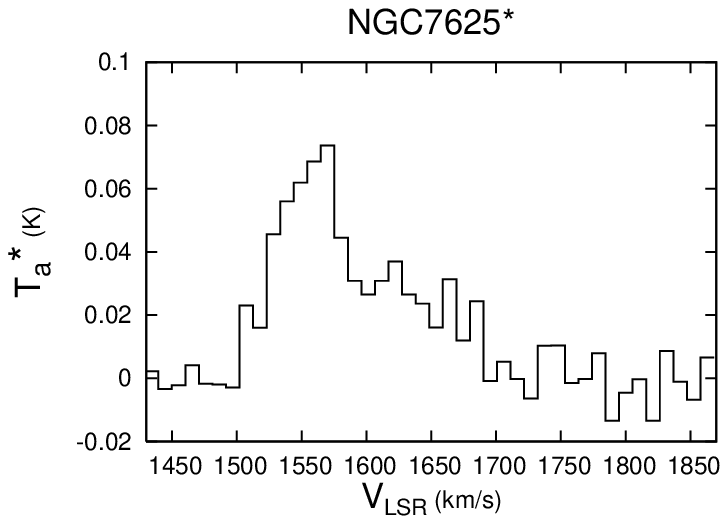}\\
     \end{tabular}
     \caption{$\atom{CO}{}{12}$ (\textit{J}\ =\
 3--2) spectra of the galaxy centers observed at ASTE.  Spectra
   with an asterisk are those observed at positions as explained in
   table \ref{astesample}. }\label{spectra}
     \end{center}
\end{figure}

The integrated intensity $\mathrm{I_{CO}}$(\textit{J}\ =\ 3--2)$ =\int \mathrm{T_{mb}d}v$ can be converted to
surface gas density using the conversion factor $X_\mathrm{CO}$ such
that $\Sigma \propto X_\mathrm{CO}\mathrm{I_{CO}}$, assuming that $\atom{CO}{}{12}$ (\textit{J}\ =\
 3--2) uniquely traces dense gas.  However, gas temperature can
 contribute to $\atom{CO}{}{12}$ (\textit{J}\ =\
 3--2) emission and will introduce complications.  In order to
 circumvent this difficulty, we use
$\mathrm{I_{CO}}$ (\textit{J}\ =\ 3--2)  hereafter, regardless of what
it implies physically.
  Errors for the values listed in table \ref{astesample} were
calculated using 
\begin{equation}
\delta \mathrm{I_{CO}}= \sigma \sqrt{\Delta V_{CO} \delta V} \quad [\mathrm{K\ km\ s^{-1}}]
\label{eqerror}
\end{equation}
where $\sigma$ is the r.m.s. noise in $\mathrm{T_{mb}}$, $\Delta V_{CO}$ the full line width,
and  $\delta V$ the velocity resolution (15 or 20 $\mathrm{km\ s^{-1}}$).  The obtained spectra are presented in figure \ref{spectra}.

All galaxies were detected except for NGC\ 520, but we attribute this to
NGC\ 520's line width of 500$\mathrm{km\ s^{-1}}$ \citet{solomon92}.

\subsection{SFR}
The SFR was derived for the sample galaxies where possible, using narrow
band H$\alpha$ imaging data from \citet{Young96}.   Flux within the
ASTE beam was calculated using task ``apphot'' on the FITS images using
IRAF software.  The SFR surface density $\Sigma_\mathrm{SFR}$ was then calculated using the
formulation by \citet{K98}, and corrected for inclination by a
factor $\cos i$.

\[
\Sigma_\mathrm{SFR} \quad (\MO\ \mathrm{pc}^{-2}\ \mathrm{yr}^{-1})
\]
\begin{equation}
=7.9\times 10^{-42} \frac{L(H\alpha)}{\mathrm{S}}\cos i \quad (\mathrm{ergs} \ \mathrm{s}^{-1} \mathrm{pc}^{-2}) 
\label{kennicuttha}
\end{equation}
where $S$ is the projected area of the observing beam.  The SFR derived in this way was then corrected for
internal extinction into $\Sigma_\mathrm{SFR}^{corr}$ when possible, by dust using $E(B-V)$ magnitudes derived
from $H \alpha /H \beta$ ratio given in \citet{Ho97}. 
Combined with equation \ref{kennicuttha}, we obtain the extinction
corrected SFR $\Sigma_\mathrm{SFR}^{corr}$. 

In order to compensate for the lack of data, we compiled $\atom{CO}{}{12}$ (\textit{J}\ =\
 3--2) from other sources.  Table \ref{other} lists these sources, along
 with the SFR derived as above where possible.  

\subsection{Schmidt Law}

%%%%%%%%%%%%%%%%%%%%%%%%%%%%%%%%%%%%%%%%%%%%%%%%%%%%%%%%%%%%%%%%%%%%%%%%%%%%%%%%%%%%%%%%%%%
%%%%%%%%%
\begin{table}
\caption{Sample Galaxies.}\label{other}
\begin{center}
\begin{tabular}{lllll}  \hline \hline
 Galaxy & Ref.  &  $\mathrm{I_{CO}^{3-2}}$ & $\log \Sigma_\mathrm{SFR}$ & $\log \Sigma_\mathrm{SFR}^{corr}$ \\
  &     &  (K km/s)  & $\MO\ \mathrm{pc}^{-2}\ \mathrm{yr}^{-1}$& \\ \hline
 NGC\ 891  & 1 & 24$\pm$2      & -9.01 & -7.64 \\
 NGC\ 2146 & 1 & 193           & -6.88 & -5.76 \\
 NGC\ 2276 & 2 & 12.2$\pm$0.2  & -7.21 & -6.39 \\
 NGC\ 2903 & 1 & 63$\pm$2      & -6.78 & -6.37 \\
 NGC\ 3079 & 2 & 183$\pm$3.2   & -8.14 & -6.06 \\
 NGC\ 3351 & 1 & 28            & -6.73 & -6.33 \\
 NGC\ 3627 & 1 & 8.8$\pm$1     & -7.23 & -6.63 \\
 NGC\ 4088 & 1 & 26$\pm$1      & -7.73 & -6.89 \\
 NGC\ 4102 & 1 & 31            & -6.97 & -6.04 \\
 NGC\ 5907 & 1 & 6$\pm$0.7     & -9.43 & -8.18 \\
 NGC\ 6946 & 1 & 46$\pm$2      & -7.00 & -5.92 \\
 NGC\ 7331 & 1 & 17.5$\pm$1.2  & -7.49 & -7.27 \\
 NGC\ 7541 & 1 & 7$\pm$1       & -7.96 & ---   \\ \hline \hline
\end{tabular}
\end{center}
Col.(1): Galaxy name.  Col.(2): References.  1 refers to
 \citet{mauersberger}, 2 refers to \citet{narayanan}.  Both were
 observed at the HHT, with an angular resolution of 22$^{\prime \prime}$, same as ASTE.  Col.(3):
 Integrated intensity of $\atom{CO}{}{12}$ (\textit{J}\ =\
 3--2) line, in main beam temperature units.  Col.(4)(5): SFR and
 extinction corrected SFR, respectively.
\end{table}
%%%%%%%%
%%%%%%%%%%%%%%%%%%%%%%%%%%%%%%%%%%%%%%%%%%%%%%%%%%%%%%%%%%%%%%%%%%%%%%%%%%%%%%%%%%%%%%

\begin{figure}[htbp]
\includegraphics[width=9cm,height=9cm,keepaspectratio]{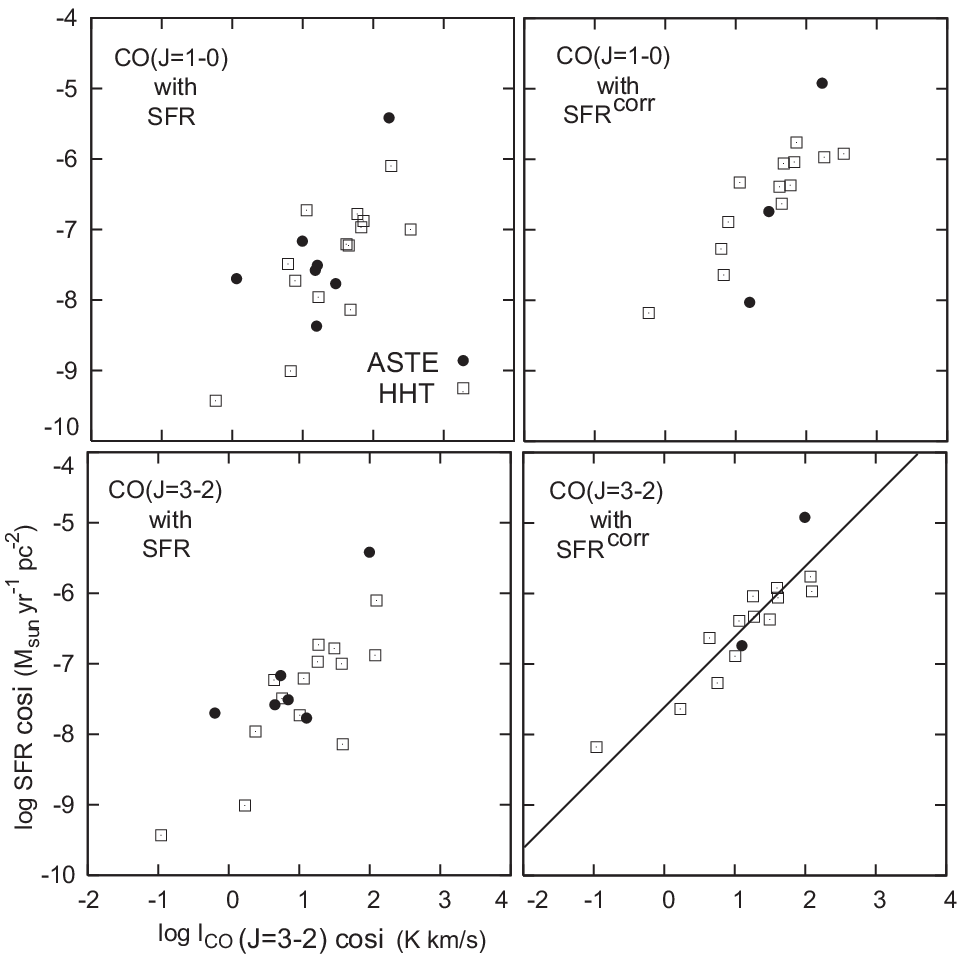} \\
  \caption{The obtained Schmidt law for the sample galaxies, using combinations of gas ($\atom{CO}{}{12}$ (\textit{J}\ =\ 1--0) 
and $\atom{CO}{}{12}$ (\textit{J}\ =\ 3--2)), and SFR with and without internal extinction correction. 
 The lower right hand panel is best correlated, with $N=1.0$ shown as the best fit line.} \label{co32a}
\end{figure}

Figure \ref{co32a} shows the obtained Schmidt law between extinction
corrected H$\alpha$ and $\atom{CO}{}{12}$ (\textit{J}\ =\
 3--2).  For comparison, we also show the relation between
 $\atom{CO}{}{12}$ (\textit{J}\ =\
 1--0) and H$\alpha$.  $\atom{CO}{}{12}$ (\textit{J}\ =\
 1--0) data are taken from similar resolution (16$^{\prime \prime}$ or
 22$^{\prime \prime}$) surveys by Komugi et al. (in preparation), \citet{braine}, and \citet{nishiyama}. 
 Apparently, a combination of extinction corrected
 SFR and dense gas gives a better correlation.  Table \ref{fits} gives
 the Schmidt law index $N$ and correlation coefficient.  In all cases, the index $N$ of the Schmidt law is
 found to be effectively 1, consistent with previous studies which use
 total luminosity in comparing the two values representing gas and SFR
 \citep{gao, yao, boker}, even though we use surface
 densities which are more indicative of the intrinsic properties of the
 samples.

%%%%%%%%%%%%%%%%%%%%%%%%%%%%%%%%%%%%%%%%%%%%%%%%%%%%%%%%%%%%%%%%%%%%%%%%%%%%%%%%%%%%%%%%%%%
%%%%%%%%%
\begin{table}
\caption{Least Squares fit}\label{fits}
\begin{center}
\begin{tabular}{lcll}  \hline \hline
Gas tracer & Ext. Corr. & $N$ & $r^2$ \\ \hline
$\atom{CO}{}{12}$ (\textit{J}\ =\ 3--2) & Yes & 0.93$\pm$0.12 & 0.91 \\ 
&     & 0.86$\pm$0.19 & 0.78 \\
$\atom{CO}{}{12}$ (\textit{J}\ =\ 1--0) & Yes & 1.05$\pm$0.19 & 0.83 \\
&     & 1.00$\pm$0.30 & 0.41 \\
$\atom{CO}{}{12}$ (\textit{J}\ =\ 3--2) & No  & 0.92$\pm$0.17 & 0.79 \\
&¡¡   & 0.60$\pm$0.21 & 0.57 \\
$\atom{CO}{}{12}$ (\textit{J}\ =\ 1--0) & No  & 0.93$\pm$0.22 & 0.70 \\ 
&¡¡¡¡ & 0.48$\pm$0.25 & 0.41 \\
\hline \hline
\end{tabular}
\end{center}
Col.(1): Tracer used for dense gas.  Col.(2): ``No'' for no H$\alpha$
 extinction correction, ``Yes'' for correction applied as explained in
 text.  Col.(3)(4): Schmidt law index $N$ and correlation coefficient, from a least
 squares fitting.  For all gas tracers, the lower row is for no inclination correction.
\end{table}

%%%%%%%%
%%%%%%%%%%%%%%%%%%%%%%%%%%%%%%%%%%%%%%%%%%%%%%%%%%%%%%%%%%%%%%%%%%%%%%%%%%%%%%%%%%%%%%
%%%%%%%%%%%%%%%%%%%%%%%%%%%%%%%%%%%%%%%%%%%%%%%%%%%%%%%%%%%%%%%%%%%%%%%%%%%%%%%%%%%%%%%
%%%%%%%%%%%%%%

\begin{figure}[htb]
\includegraphics[width=9cm,height=12.5cm,keepaspectratio]{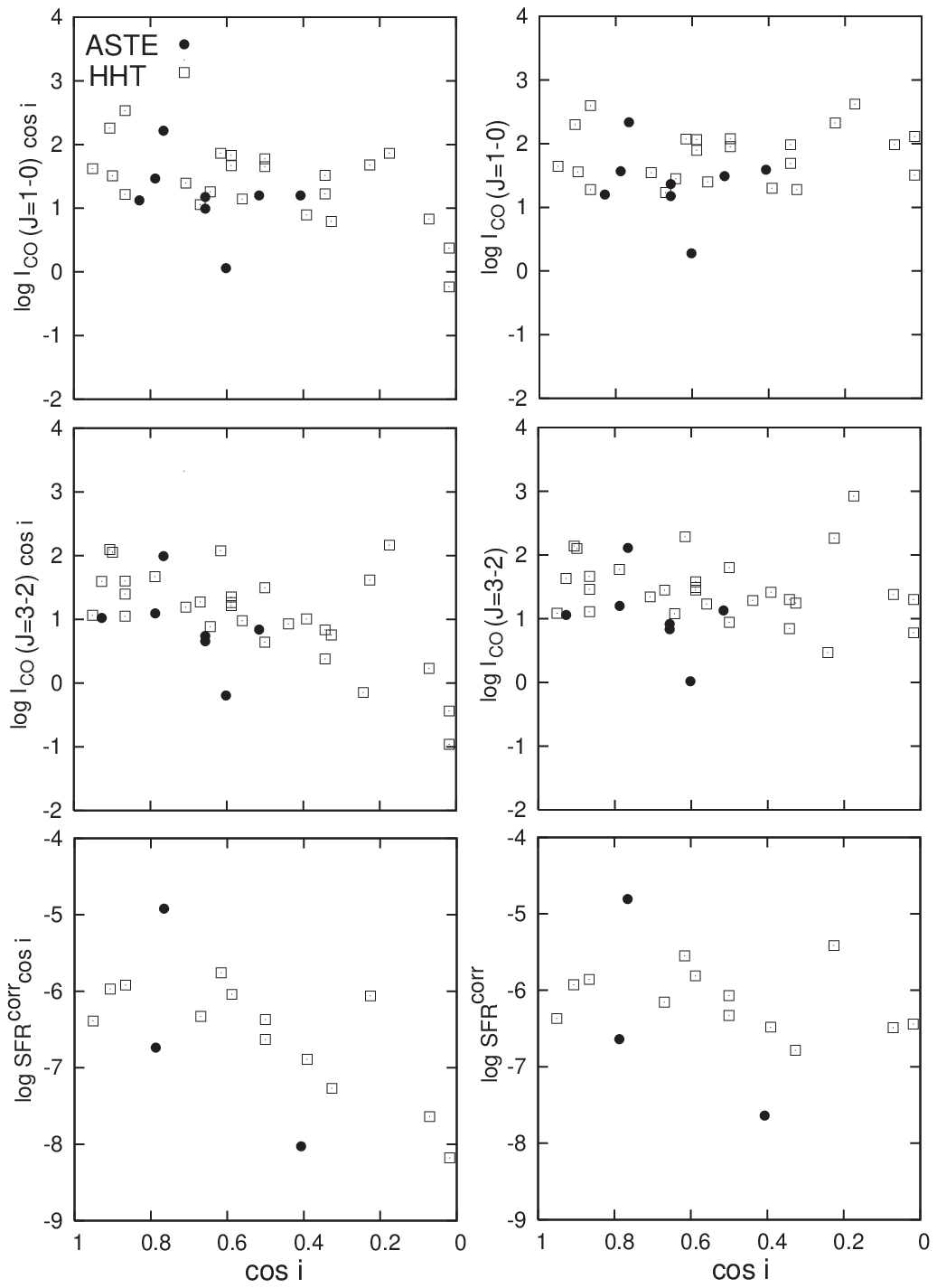} \\
  \caption{Inclination versus surface gas density and SFR density,
 corrected for (left) and uncorrected for inclination (right).  Notice
 that figures on the left show decreasing density with inclination,
 whereas figures on the right show no trend.} \label{inclination}
\end{figure}
 
\begin{figure}[htb]
 \includegraphics[width=9cm,height=8cm,keepaspectratio]{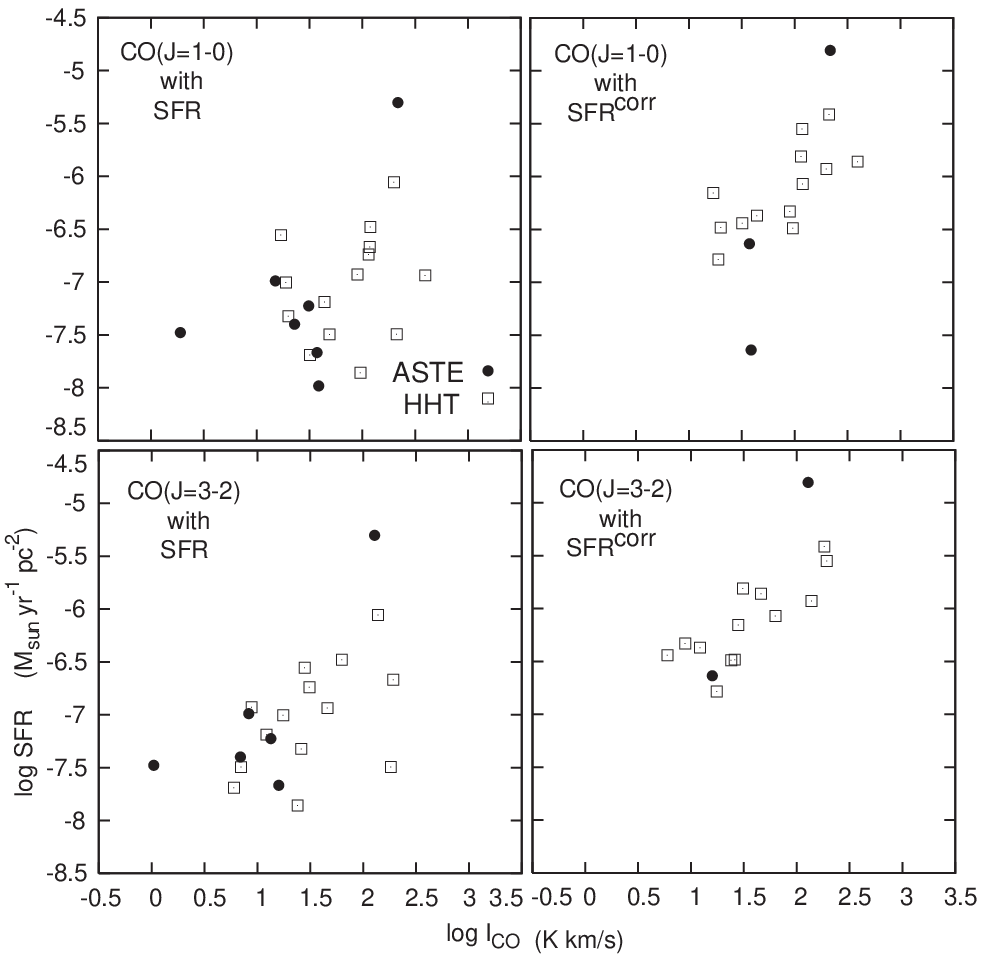} \\
  \caption{Same as figure 2, but uncorrected for inclination.
 $\atom{CO}{}{12}$ (\textit{J}\ =\ 3--2) with corrected SFR (correlation
 coefficient 0.78, N=) still shows a better
 correlation compared to $\atom{CO}{}{12}$ (\textit{J}\ =\ 1--0)
 (correlation coefficient 0.41).} \label{co32c}
\end{figure}

%%%%%%%%%%%%
%%%%%%%%%%%%%%%%%%%%%%%%%%%%%%%%%%%%%%%%%%%%%%%%%%%%%%%%%%%%%%%%%%%%%%%%%%%%%%%%%%%%%%%

\subsection{Inclination Correction}
An effect we must consider when using surface averaged values is the
inclination of the target galaxy, where the surface density is derived
by multiplying the luminosity by $\cos i$.  However, inclination is
generally defined from its global morphology, and hence is not always
indicative of the disk in its central kpc, as in this case.
 The central region
may be thick or clumpy, so that corrections for inclination are not
necessary in the first place.  To see if a factor of $\cos i$ is
justified, we plot the inclination corrected gas densities and SFR vs.
$\cos i$ in the left hand side of figure
\ref{inclination}.  Apparently, the values decrease with increasing
inclination (i.e., as they become edge-on), a trend which should not be seen if the correction is
correct, because inclination of galaxies is a meaningless parameter regarding
its intrinsic properties, and the average face-on surface density should be
independent of inclination.  The decreasing trend is roughly the same for
$\atom{CO}{}{12}$ (\textit{J}\
=\ 1--0), $\atom{CO}{}{12}$ (\textit{J}\
=\ 3--2) and SFR, which suggests that this
trend is not due to opacity of the emission lines, but more of a
geometric quality, i.e., that the structure inside the 22$^{\prime
\prime}$ beam is more clumpy than disky.  The
resulting Schmidt law will be biased towards $N=1$, because the gas and SFR
densities decrease equally along a slope of 1.  The right hand side
of figure \ref{inclination}, is the same for values uncorrected for inclination.  We
do not see any trend with inclination, which is physically more
plausible. 

From these views, we conclude that a correction for inclination of $\cos
i$ is not completely justified; figure \ref{co32c} show the Schmidt law for all
values uncorrected for inclination.  The variance of the residuals for a
best fit is given in table \ref{fits}, and in this case also we see that
the use of $\atom{CO}{}{12}$ (\textit{J}\
=\ 3--2) and extinction corrected H$\alpha$
improves the fit greatly.

\section{Discussion}

We have shown for the first time that $\atom{CO}{}{12}$ (\textit{J}\
=\ 3--2) has a striking correlation with H$\alpha$ derived SFR.
Assuming that $\atom{CO}{}{12}$ (\textit{J}\
=\ 3--2) indeed traces dense gas, this implies that H$\alpha$ may be a valid SFR tracer even in dense
and dusty regions.  Dense gas $\atom{CO}{}{12}$ (\textit{J}\
=\ 3--2) also correlates better with recent
star formation compared to $\atom{CO}{}{12}$ (\textit{J}\ =\ 1--0).

Interpreting this result in a qualitative manner is easy.  Assuming that
star formation occurs where gas density exceeds a certain value, we can
expect that $\atom{CO}{}{12}$ (\textit{J}\
=\ 3--2) is more spatially and temporally connected to star formation compared
to $\atom{CO}{}{12}$ (\textit{J}\ =\ 1--0).  By using H$\alpha$ as a SFR tracer, the spatial connection
(resolution) and temporal connection (traces SF over $10^6$ years) are
even more improved.  This improvement should show up in terms of the
Schmidt law.  

Using H$\alpha$, we have circumvented the need to use
global gas and SFR in expressing the Schmidt law, which is known to introduce a size effect.  
Larger galaxies tend to be more luminous in any wavelength:
 therefore producing a correlation between gas and SFR without any other
 physical reasons (see \cite{stark86}).  The resolution of H$\alpha$ imaging allowed us to meaningfully
derive surface density values, whereas using the IRAS data may force us to compare spatially decoupled gas and SFR within the
large beam.

The Schmidt law index $N$, however, should be
treated with care.  The correction for extinction can be applied only to
detectable emission, and not the intrinsic total H$\alpha$
emission.  This is not unique to recombination lines, but to all star
formation tracers where extinction plays a role.  In that light, The
observed Schmidt law index is always an underestimate, where in our case
 the effect may be strong because our study focuses on the central
 regions of dusty galaxies.  This is in concord with the literature,
 where the widely accepted value of
1.4 (e.g., \cite{K98}) is considerably higher than our derived value of
$\sim$ 1.0 \\[1.0cm]

The authors thank Judy S. Young, for kindly providing us with narrow
band H$\alpha$
images.  S.K., S.O., and F.E. were financially
supported by a Research Fellowship from the Japan Society for the Promotion
of Science for Young Scientists.
 A part of this study was financially supported 
by the MEXT Grant-in-Aid for Scientific Research on
Priority Areas No. 15071202.
This paper has made use of the NASA/IPAC Extragalactic
Database (NED), which is operated by the Jet Propulsion Laboratory,
Caltech, under contract with the National Aeronautics and Space Administration.

\end{document}